\newcommand{\bea}{\begin{eqnarray}}
\newcommand{\eea}{\end{eqnarray}}
\newcommand{\bear}{\begin{eqnarray*}}
\newcommand{\eear}{\end{eqnarray*}}

\documentstyle[aps,preprint]{revtex}
\begin{document}

\draft

\title
{EXACT SOLUTION OF A VERTEX MODEL WITH UNLIMITED NUMBER OF STATES 
PER BOND}

\author{
F. C. Alcaraz$^1$ and R. Z. Bariev$^{1,2}$}

\address{$^1$Departamento de F\'{\i}sica, 
Universidade Federal de S\~ao Carlos, 13565-905, S\~ao Carlos, SP
Brazil}

\address{$^2$The Kazan Physico-Technical Institute of the Russian Academy of Sciences, 
Kazan 420029, Russia}

\maketitle

\begin{abstract}
The exact solution is obtained for the eigenvalues and eigenvectors of 
the row-to-row transfer matrix of a two-dimensional vertex model with
unlimited number of states per bond. This model is a classical 
counterpart 
of a quantum spin chain  with an unlimited value of spin. 
This quantum chain is studied using general 
predictions of conformal field theory. The long-distance behaviour 
of some 
ground-state correlation functions is derived from a finite-size
analysis of the gapless excitations. 

{\bf  Published in J.Phys.A: {\bf 32} L25 (1999)}
\end{abstract}


\narrowtext
Most of exactly solved models in two dimensions are known to be related to the 
six-vertex model solved by Lieb~\cite{1}. This model exhibits an infinite 
number of conserved integrals commuting with each other. The generator 
of these integrals of motion is the row-to-row transfer matrix. The 
conserved quantities are quantum Hamiltonians describing different 
one-dimensional systems, in particular one of these describes the XXZ chain 
[2].
There exist also exact solutions for a variety of generalizations 
of this model. In particular lattice models where the link variables 
may assume $q>2$ different states were solved using the Bethe-anstaz 
method [3-12].

In this letter we present an exactly integrable  vertex model 
in which the number of different 
states  of a given link is not limited, and  may be considered as a
generalization of the six-vertex model for the case where on each link  
of the 
lattice we may have an unlimited number of lines. This model is the classical
counterpart of a quantum spin chain  with unlimited value of the spin.
The vertex model may be formulated as follows. We consider a square 
lattice with $M$ rows and $N$ columns and with periodic boundary conditions.
At the bonds connecting adjacent pair of lattice sites we attach a variable
 $\alpha =0, 1, 2 ,\ldots$, representing the number of lines (or particles) 
in the bond. The links $(\alpha, \beta, \gamma, \delta = 0,1,2,\ldots)$ 
connected with a given site produce a vertex configuration, with 
Boltzmann weight $R_{\gamma,\delta}^{\alpha,\beta}$ as in figure 1. The 
model we consider obey a type of "ice condition" [1], where the 
Boltzmann weights are chosen such that the number of lines is 
conserved as we go from the direction $(\alpha,\beta)$ to $(\gamma,\delta)$, 
i. e., $\alpha + \beta = \gamma + \delta$. The non-zero Boltzmann weights are 
given by

\bea
R_{0,0}^{0,0} &=& a,\;\;  
R_{0,1}^{0,1} = R_{1,0}^{1,0} = b,\nonumber\\
R_{1,0}^{0,1} &=& R_{0,1}^{1,0} = c,
\eea

and
\bea
R_{1,1}^{0,2} &=&R_{0,2}^{1,1} =  R_{n,m+1}^{m,n+1} = b, \;\; 
R_{n,0}^{n,0} = b^n/a^{n-1},\nonumber\\
R_{n+l,m}^{m+l,n} &=& \frac{c^2}{a}\left(\frac{b}{a}\right)^l,\;\;
R_{l,m}^{m+l,0} = R_{m+l,0}^{l,m} = c\left(\frac{b}{a}\right)^l,
\eea
where $ m,n > 0 $, $ l \geq 0$ and $a,b,c$ are arbitrary parameters. For 
convenience we use the Baxter's parametrization [3]
\bea
a = \sinh\frac{1}{2}(\lambda - u), \;\; b = \sinh\frac{1}{2}
(\lambda + u), \;\; c = \sinh \lambda,
\eea
where $u$ and $\lambda$ are the spectral and crossing parameters, 
respectively. If we exclude in (1)-(2) the configurations where 
$\alpha,\beta,\gamma,\delta >1$ we obtain the standard six-vertex 
model [3].

To calculate the partition function of the system we use the row-to-row
transfer matrix $T$, with elements $T_{\{\alpha\},\{\alpha'\}}$,  given 
by the product of the Boltzmann weights 
connecting two sucessive rows $\{\alpha_1,\alpha_2,\ldots,\alpha_N\}$ and 
$\{\alpha_1',\alpha_2',\ldots,\alpha_N'\}$
\bea
T(u)_{\{\alpha\},\{\alpha'\}}
 = \sum_{\beta_1,\beta_2,\ldots,\beta_N = 1}^{\infty} 
R_{\alpha_1',\beta_N}^{\alpha_1 ,\beta_1}(u)
R_{\alpha_2',\beta_1}^{\alpha_2,\beta_2}(u) \ldots
R_{\alpha_N',\beta_{N-1}}^{\alpha_N,\beta_N}(u).
\eea
 The above transfer matrices with different values of the
spectral parameter commute among themselves, i. e.,
\bea
T(v)T(u) = T(u)T(v).
\eea
A sufficient condition for this commutativity  is the existence of 
$X_{\gamma,\delta}^{\alpha,\beta}(u,u')$ satisfying the Yang-Baxter 
equations [3]
\bea
\sum_{\gamma,\delta'',\mu''} 
R_{\alpha,\delta}^{\gamma,\delta''}(u)R_{\gamma,\mu}^{\beta,\mu''}(u')
X_{\delta'', \mu''}^{\mu',\delta'}(u,u') = 
\sum_{\gamma,\delta'',\mu''}X_{\delta, \mu}^{\mu'',\delta''}(u,u')
R_{\alpha,\delta''}^{\gamma,\delta'}(u')
R_{\gamma,\mu''}^{\beta,\mu'}(u).
\eea
Although we did not succeed in finding $X_{\gamma,\delta}^{\alpha,\beta}(u,u')$ 
 we have checked equation (5) directly.
We want to solve the eigenvalue equation of  the transfer matrix 
\bea
T\Psi = \Lambda\Psi.
\eea
An obvious way to describe the state of a row is to specify the positions of 
individual particles (lines).
Let
\bea
\Phi(x_1,x_2,\ldots ,x_n)
\eea
be the amplitude corresponding to the state of a row with particles 
(lines) at the
bonds  $x_1,x_2,\ldots ,x_n $. 
Unlike the ordinary six-vertex model,  some of $x_i$
may coincide
\bea 
x_1 \leq x_2 \leq \ldots \leq x_n 
\eea
A possible way to represent the model under consideration 
is to draw as much lines on
an edge as the value of the variable $\alpha$  
(number of particles) of this edge. 
An example of  arrangement of lines is 
shown in figure 2. From the definition of our model (1-2) and 
Fig. 2 it is clear 
that as in the ordinary six-vertex model~\cite{1}  the lines link together 
by forming   
continuous noncrossing paths through the lattice. If one starts by following 
a path upwards or to the right, then one will always be travelling in one or 
other of these two directions, never downwards or to the left. The cyclic 
boundary 
conditions ensure
that a path never ends. A typical configuration of lines in adjacent rows is
shown in Fig. 3. Consequently
the total number of particles (lines) on arrow is a conserved quantity 
and  
our model can be treated similarly as in [1,3].
 However in our case the calculation is more complicated. 
We write the amplitude in (8) in the form of the Bethe ansatz [3,4]
\bea
\Phi(x_1,\ldots ,x_n) = \sum_P A_{P_1\ldots P_n}\exp(ik_{P_1}x_1+\ldots 
+ik_{P_n}x_n),
\eea
where the sum is over all the permutations $P$ of $(1,2,\ldots,n)$ and 
${k_{i}} (i=1,2,...,n)$ are unknown quasimomenta.
The eigenvalues $\Lambda_{\{k\}}$  corresponding to the eigenvector 
$\Psi_{\{k\}}$ with amplitudes (10) are given by 
\bea
\Lambda_{\{k\}} = a^N\prod_{j=1}^n L(k_j) + b^N\prod_{j=1}^n N(k_j),
\eea
where
 $L(k)$ and $N(k)$ can be calculated by considering 
 a single particle 
$n=1$. In our case we have
\bea
L(k) &=& [ab + (c^2-b^2)e^{ik}]/[a^2 - abe^{ik}], \nonumber\\
N(k) &=&e^{ik}[1 - c^2(a^2-abe^{ik})^{-1}].
\eea
Considering the eigenvalue  equations at the boundaries of 
the inequalites (9) after a long, but straightforward calculation,
we obtain the conditions for the coefficients $A_{P_1...P_n}$,
\bea
A_{\ldots P_1P_2\ldots } &=& -(1 - 2\Delta e^{ik_{P_1}} + e^{ik_{P_1}+ik_{P_2}})
\nonumber\\ &\cdot&(1 - 2\Delta e^{ik_{P_2}} + e^{ik_{P_1}+ik_{P_2}})^{-1}
e^{ik_{P_2}- ik_{P_1}}A_{\ldots P_2P_1\ldots },\\
A_{P_1\ldots P_n}&=& A_{P_2\ldots P_n P_1}\exp(ik_{P_1}N),
\eea
where
\bea
\Delta &=&(a^2 + b^2 - c^2)/2ab = -\cosh\lambda .\nonumber
\eea

The Bethe-ansatz equations that come from (13)-(15) and fix the values of 
$\{k\}$ are simplified by using the standard parametrization of the 
six-vertex model [3]
\bea
e^{ik_j}& =& \frac{\sinh\frac{1}{2}(\lambda -u_j)}{\sinh\frac{1}{2}
(\lambda +u_j)},\nonumber\\
\eea
 and are given by 
\bea
\left[\frac{\sinh\frac{1}{2}(\lambda +u_j)}{\sinh\frac{1}{2}(\lambda -u_j)}
\right]^{N+n} &=&(-1)^{n-1}e^{-iP}\prod_{l=1}^n
\frac{\sinh[\lambda+\frac{1}{2}(u_j -u_l)}{\sinh[\lambda -\frac{1}{2}(u_j -u_l)}
\eea
where 
\bea
P = \sum_{l=1}^n k_l \; (\mbox{mod} 2\pi) \nonumber
\eea
is the associated momentum of the eigenstate $\Psi_{\{k\}}$. The eigenvalue 
(11) is now given by 
\bea
\Lambda(u)_{\{k\}} = a^N\prod_{j=1}^{n}\tilde L(u_j) + b^N(\frac{b}{a})^n\
exp(iP)
\prod_{j=1}^{n}\tilde M(u_j)
\eea
where
\bea
\tilde L(u_j)=\frac{\sinh[\lambda+\frac{1}{2}(u-u_j)]}{\sinh\frac{1}{2}(u_j-u)}
\nonumber\\
\tilde M(u_j) = 
\frac{\sinh[\lambda-\frac{1}{2}(u-u_j)]}{\sinh\frac{1}{2}(u-u_j)}.
\eea
Equations (10)-(18) completely describe the eigenvectors and eigenvalues of 
the row-to-row transfer matrix of the model with Boltzmann weights (1)-(2).
 From the comparison of these equations with the 
corresponding equations of  the 
six-vertex model it is clear that  the eigenvalues of the transfer
matrix of the model under consideration up to the multiplicative factor $a^n$
coincide with the eigenvalues  of the transfer matrix of a six-vertex
model with $N+n$ columns but with Boltzmann weights modified along a vertical 
"seam" in the lattice [13,14]. 
The boundary angle, that modifies the Boltzmann weights along the seam is 
equal to the momentum $P$  of the state.

Let us consider the thermodynamic limit where $n \rightarrow \infty$ and  
$ N \rightarrow  \infty$, 
but the density of particles (or density of lines) $\rho =n/N$
is kept fixed. The equations (16)-(17) can be rewritten in terms of 
integral equations.
In the region $ -1 \leq \Delta < 1$, $\lambda = i\mu$, $0 \leq \mu < \pi $ 
we have
\bea
2\pi\rho(u) +\int_{-u_0}^{u_0}\frac{\sin(2\mu)\rho(u')}
{\cosh(u-u')-\cos(2\mu)}du' = \frac{\sin\mu}{\cosh u -\cos\mu}(1+\rho),
\eea
\bea
\int_{-u_0}^{u_0}\rho(u)du=\rho,
\eea
\bea
\frac{1}{N}\ln[\Lambda(u)] =-\rho\ln a &+& (1 +\rho)\max\lbrace\ln a
 + \int_{-u_0}^{u_0}
\ln[\tilde L(u)]\rho(u)du; \nonumber\\
\ln b  &+& \frac{iP}{N} + \int_{-u_0}^{u_0}\ln[\tilde M(u)]\rho(u)du\rbrace.
\eea
Since the maximum eigenvalue has momentum zero it is interesting
to compare equations (19)-(20) with the corresponding ones in the 
six-vertex model [1]. This comparison enables us to express the free energy 
$f(\rho)$ of our model, at a given density $\rho$, in terms of
the free energy $f^{6v}(\rho)$ of the six-vertex model, i.e.

\bea
f(\rho) =\beta^{-1}\rho\ln a + (1 +\rho)f^{6v}(\frac{\rho}{1+\rho}).
\eea

We can also obtain an exactly integrable quantum Hamiltonian, 
related with our vertex model 
by taking the logarithmic derivative of the
row-to-row transfer matrix (4) at $u = -\lambda$.
This gives us a new integrable quantum model with Hamiltonian
\bea
H &=& -N\cosh\lambda + 2\cosh\lambda\sum_{j=1}^L\sum_{m=1}^{\infty}
(m-1)E_j^{m,m}
\nonumber\\
&+&\sum_{j=1}^L\sum_{m,n=1}^{\infty}(E_j^{m,m+1}E_{j+1}^{n+1,n}+
E_j^{m+1,m}E_{j+1}^{n,n+1})
\eea
where $E_{j}^{n,m}$ are infinite-dimensional matrices with elements given by
$(E^{l,m})_{ij} = \delta_{l,i}\delta_{m,j}$.
This Hamiltonian describes the dynamics of particles on a lattice with 
no volume exclusion. It contains a hopping term (second sum) which allows 
the motion of one of the particles at a given site to its neighbouring 
sites, independently of its relative occupation and a static interaction 
(first sum) proportional to the occupation number at the site, $n_i =
\sum_{m=1}^{\infty} mE_i^{m,m}$.
The line conservation in the vertex model translates into the commutation 
of (23) with the total number of particles $n = \sum_{i=1}^L n_i$. The 
eigenenergies in a given eigensector with $n$ particles are given by 
$E_{\{k\}} = -2\sum_{j=1}^n \cos k_j$, up to an additive constant which we 
have dropped, where $\{k_j\}$ are the solutions of (16). In the thermodynamic 
limit the ground-state energy is obtained in terms of the solution 
$\rho(u)$ of (19)-(20), namely
\bea
\frac{E_0}{N} = 2\int_{-u_0}^{u_0}\lbrack\cos\mu - \frac{\sin^2\mu}
{\cosh u - \cos\mu}\rbrack\rho(u)du.
\eea
In figure 3 we show the ground-state energy per site for 
$0 < \rho \leq 1$, obtained by solving numerically (19), (20) and (24).

To study the physical properties of this quantum chain we 
calculated the long-distance behaviour of the correlation functions by 
exploring the consequences of  conformal invariance (see [15-18] and 
references therein). The critical fluctuations are governed by a 
conformal field theory with central charge $c = 1$ and as in the XXZ chain 
the critical exponents are like those of a Gaussian model [19,15]. The 
long-distance 
behavior of the density-density and of the pair-pair correlation 
functions are 
given by
\bea
\left<\rho(r)\rho(0)\right>\simeq\rho^2+A_1r^{-2}+A_2r^{-\alpha}
\cos(2\pi\rho r) 
\eea
and 
\bea
\left<\eta^+(r) \eta(0)\right> \simeq B r^{-\beta},
\eea
where $ \rho (r) = \sum_{m=1}^\infty mE_{r}^{m,m}$ and 
$\eta(r) = \sum_{m=2}^{\infty}E_r^{m-2,m}$ are the density and pair 
operators. The exponents $\alpha$ and $\beta$ describing the algebraic 
decay of the correlations are given by 
\bea
\alpha = \beta^{-1} = 2[\xi(u_0)]^2,
\eea
where the dressed charge function $\xi(u)$ is obtained by solving 
the integral equation
\bea
\xi(u) + \frac{1}{2\pi}\int_{-u_0}^{u_0}\frac{\sin(2\mu)\xi(u')}
{\cosh(u-u') - \cos(2\mu)}du' = (1+\rho).
\eea
As usual in one dimension the model does not 
 show finite off-diagonal long-range order. However,
the power decay of  the pair-pair  correlation (27) indicates the
preference of pairing of particles whenever  the exponent $\beta$ of this
correlation is smaller than that of the density-density correlation
$\alpha$ [16]. The exponent $\beta$ is plotted in figure 4 for $0 < \rho 
\leq 1$ 
and 
 some values of the parameter $\mu$.  
These results are obtained by solving  numerically the integral equations 
(19-20) and (28). We see from this figure that for all values
$-1 \leq \Delta=-\cos \mu < 1$ there exists 
a regime with dominant pair correlations. 
The analogous behaviour of correlation
functions was also  observed in the models considered in Refs. [20-21]. 
But differently 
from these models, in the present case we have the nice feature that for 
increasing values of $\rho$, the exponent $\beta$  for $0 \leq \Delta < 1$
tends toward zero and 
 there is no 
 regime where  the density-density  correlations dominate. 

It should be emphasized that the formalism exposed here for creation of new 
integrable models applies to lattice systems with higher spin like the 
Perk-Shultz
[8] and Sutherland models [10] as well as Fateev-Zamolodchikov [11] and 
Izergin-Korepin models [12].

This work was supported in part by CNPq anf FINEP-Brazil.

\begin{figure}[tbp]
\caption{
The vertex formed by the link variables $\alpha,\beta,\gamma,\delta$ and 
having Boltmann weight $R_{\gamma ,\delta}^{\alpha, \beta}$.}
\end{figure}

\begin{figure}[tbp]
\caption{
An example of a line  representation of a configuration in the 
unlimited vertex model.}
\end{figure}

\begin{figure}[tbp]
\caption{
A typical arrangement of lines of two adjacent rows.}
\end{figure}

\begin{figure}[tbp]
\caption{  
The ground-state energy per site $E_0/N$ as a function of the 
density $\rho = n/N$  and anisotropy $-1 \leq \Delta =-\cos \mu <1$, 
with (a) $\mu = \pi /10$, (b) $\mu =\pi /4$, (c) $\mu =\pi /3$, (d)
 $\mu=\pi /2$, (e) $\mu =5 \pi /6$  and (f) $\mu = 10\pi /12$.} 
\end{figure}

\begin{figure}[tbp]
\caption{
The critical exponent $\beta$ as a  function of the 
density $\rho = n/N$ , and for some values of the anisotropy 
$ \Delta =-\cos \mu $, 
with (a) $\mu = \pi/10$, (b) $\mu =\pi/6$, (c) $\mu =\pi/4$, (d) 
$\mu=\pi/2$, (e) $\mu =5\pi/6$  and (f) $\mu = 10\pi/12$.} 
\end{figure}

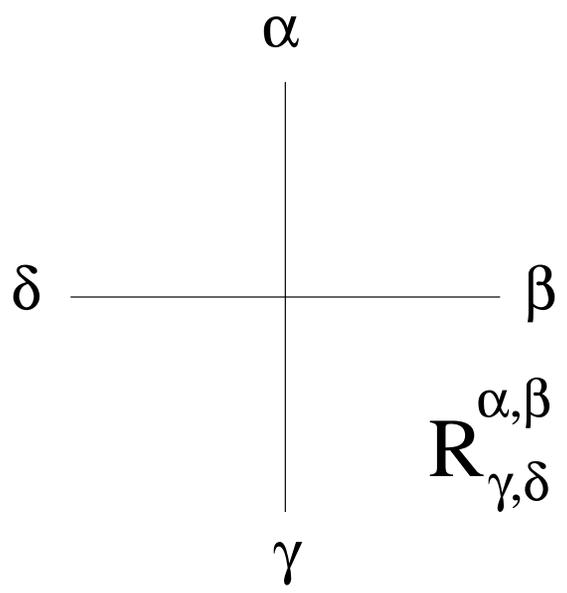
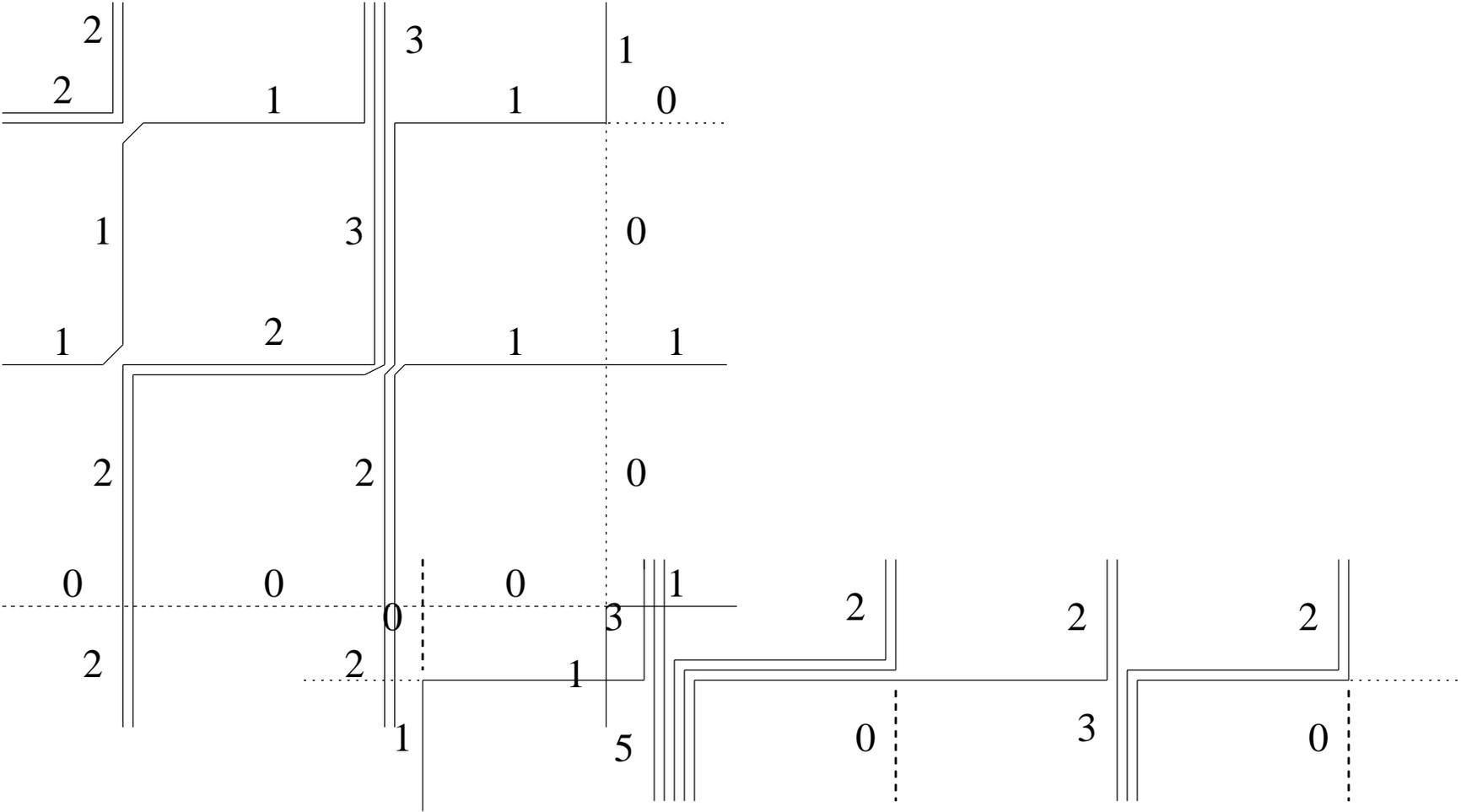
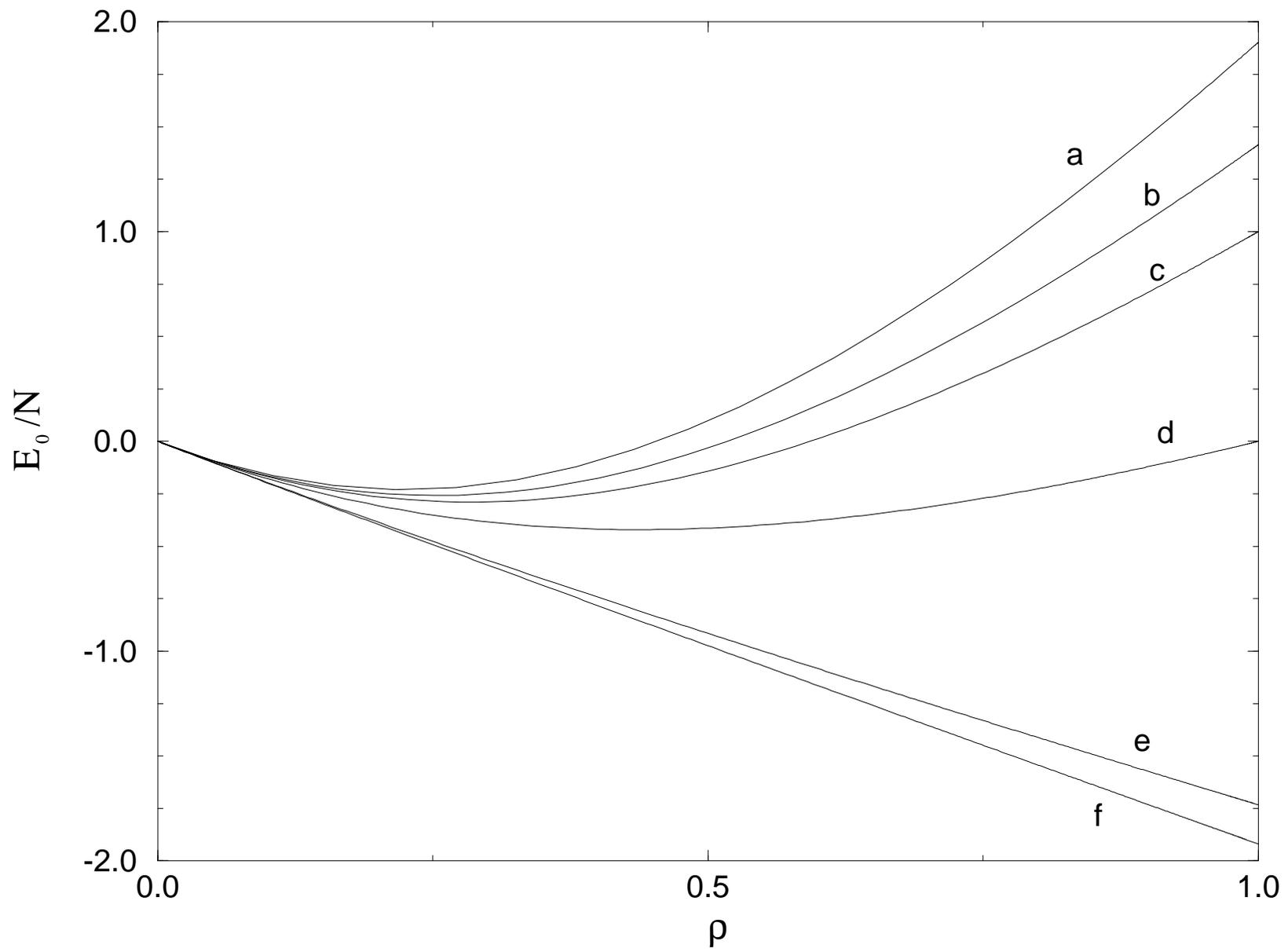
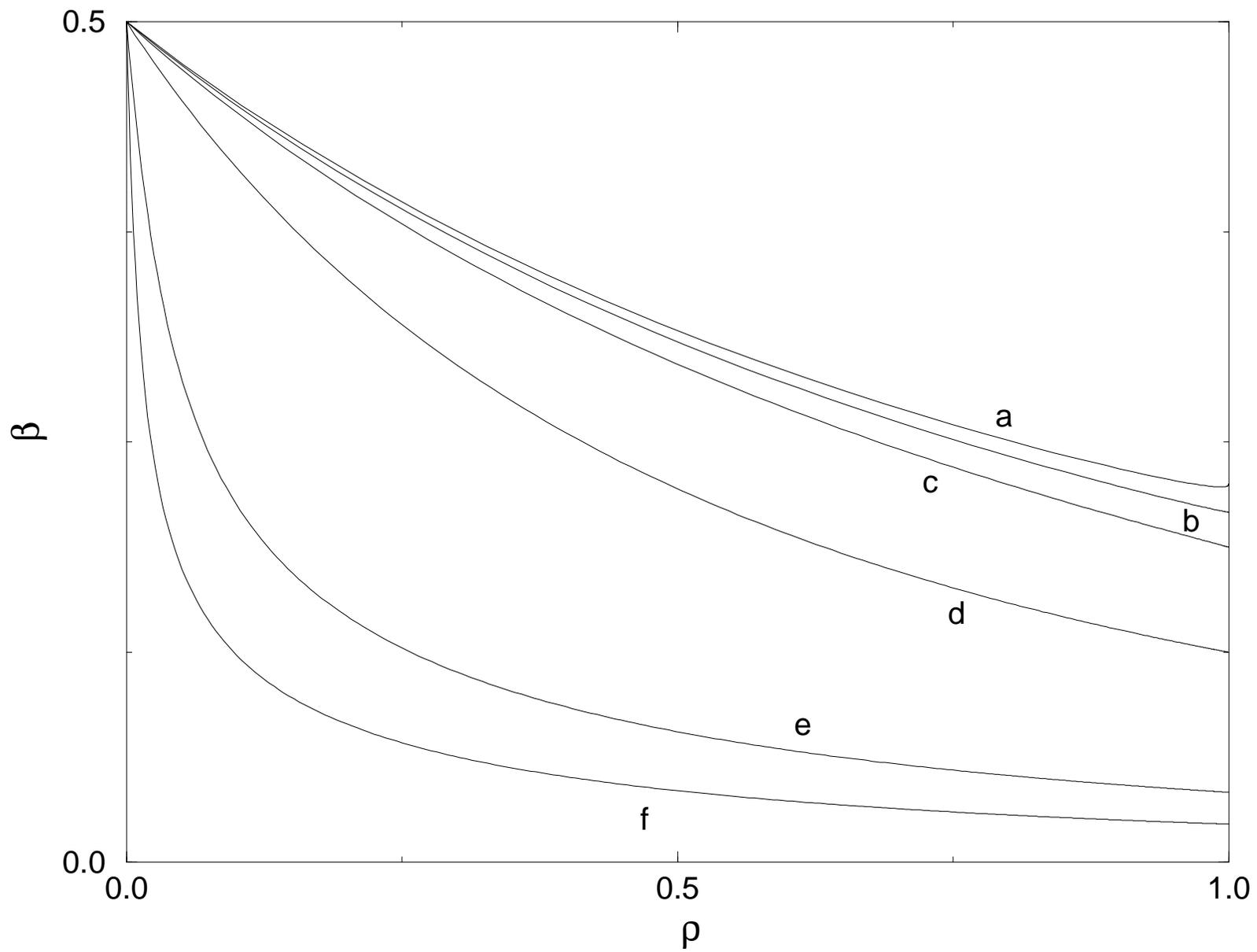
\end{document}